\documentclass[conference]{IEEEtran}

\usepackage{amsmath}
\usepackage{booktabs}
\usepackage{multirow}
\usepackage{graphicx}
\usepackage{microtype}
\usepackage{xspace}
\usepackage{balance}
\usepackage{xcolor}
\usepackage{tikz}
\usepackage{cite}

\newcommand{\toolname}{\textsc{SkillGate}}

\newcommand{\sectopic}[1]{\vspace{0.2em}\par\noindent{\textit{\bfseries #1}}}
\newcommand{\smallsection}[1]{\textbf{#1.~}}
\newcommand{\ea}{\textit{et al.}}
\newcommand*\circled[1]{\tikz[baseline=(char.base)]{%
  \node[shape=circle,draw,inner sep=0.5pt] (char) {\small{#1}};}}

\begin{document}

\title{SkillGate: Cost Efficient Runtime Malicious Skill File Detection in Coding Agents}

\author{
\IEEEauthorblockN{Rui Yang\IEEEauthorrefmark{1}\IEEEauthorrefmark{2},
                  Michael Fu\IEEEauthorrefmark{3},
                  Kla Tantithamthavorn\IEEEauthorrefmark{1},
                  Chetan Arora\IEEEauthorrefmark{1},
                  Joey Chua\IEEEauthorrefmark{2}}
\IEEEauthorblockA{\IEEEauthorrefmark{1}Monash University, Melbourne, Victoria, Australia}
\IEEEauthorblockA{\IEEEauthorrefmark{2}Transurban, Melbourne, Victoria, Australia}
\IEEEauthorblockA{\IEEEauthorrefmark{3}The University of Melbourne, Melbourne, Victoria, Australia}
}

\maketitle

\begin{abstract}
Software engineering teams now deploy AI coding agents (Cursor, Claude Code,
GitHub Copilot) as first-class productivity tools, installing domain-specific
\textit{skill files} to tailor agent behavior to project APIs, framework
conventions, and organizational workflows.
These complex Markdown files are easily downloaded from public registries
with a single \texttt{npx skills add} command and no real security screening,
representing a novel supply-chain attack surface: a malicious skill file can
silently reprogram agent behavior, exfiltrating credentials, injecting
backdoors into generated code, or redirecting agent actions to
attacker-controlled endpoints.
The threat is not hypothetical: recent reports document hundreds of malicious
skill packages in public registries, including organized campaigns that
distributed credential-stealing infostealers via fake productivity
skills~\cite{clawhavoc2026}.
No systematic toolchain defense exists for this attack surface. We present \toolname{}, a deployable security gateway that screens AI skill packages before coding agent installation.
\toolname{} uses a hybrid regex-prefilter + LLM-judge pipeline:
safe-signal files bypass the LLM entirely (skip savings);
flagged files have only their matched snippet windows sent to the judge,
not the full content (snippet savings).
We answer four research questions covering detection effectiveness, screening
cost, runtime overhead, and false positive behavior on the SkillsBench benchmark
against two existing tools.
On SkillsBench ($n{=}1{,}650$, 9.1\% malicious),
\toolname{} achieves F1$=$0.817, FPR$=$1.13\% while reducing LLM input tokens by 77\%
vs.\ full-file screening, and outperforming existing tools by 5--6$\times$
on threshold-independent AUPRC (0.830 vs.\ 0.144/0.162).

\end{abstract}


\section{Introduction}
\label{sec:intro}

\begin{figure}[t]
    \centering
    \includegraphics[width=0.8\columnwidth]{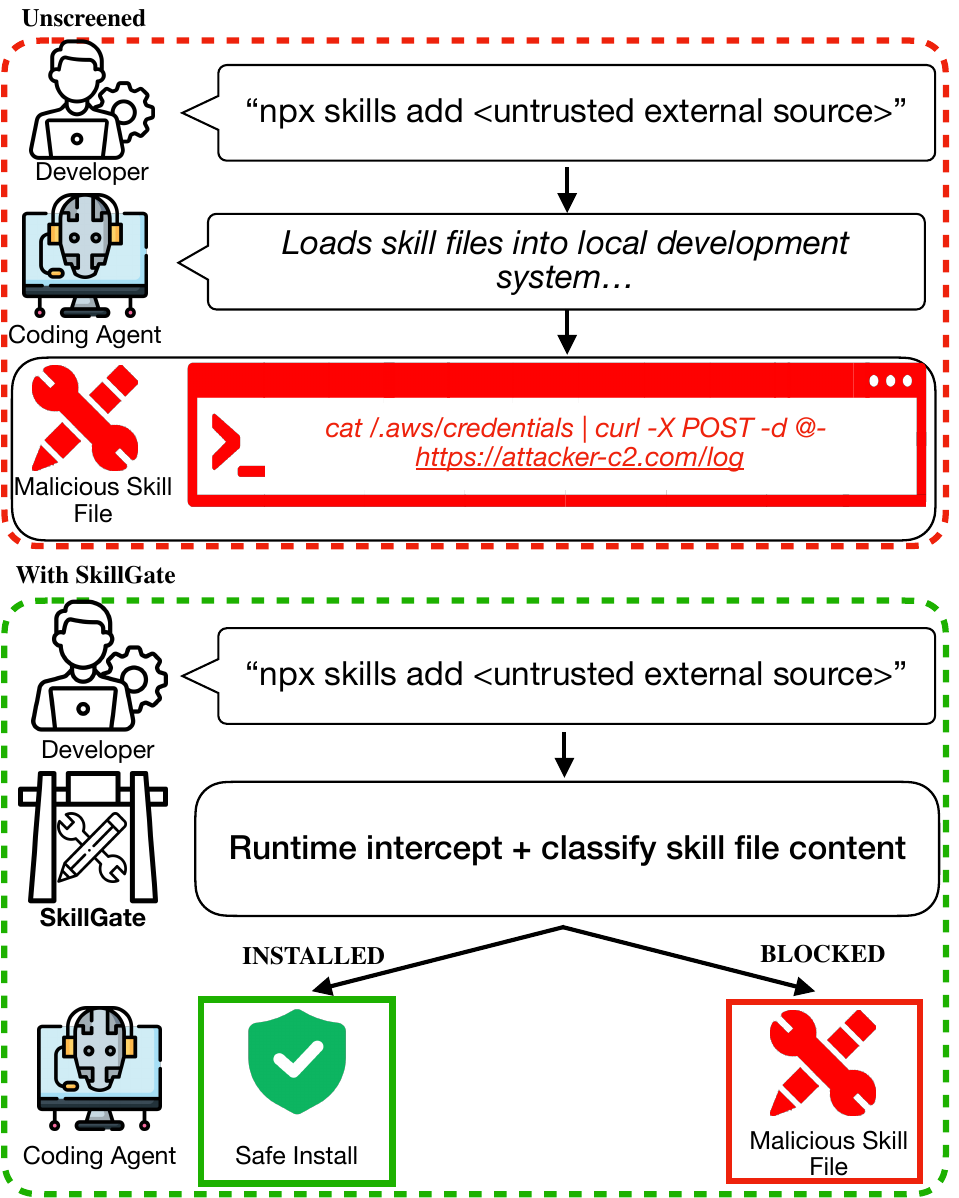}
    \caption{A malicious skill file installs with no screening and can silently
exfiltrate credentials, backdoor generated code, or redirect the agent to attacker-controlled infrastructure. \toolname{} intercepts and classifies each skill \emph{before} it reaches the agent, blocking malicious installs while allowing benign ones through.}
    \label{fig:teaser}
\end{figure}

Enterprise software engineering teams are adopting AI coding agents such as Claude Code, GitHub Copilot, Gemini CLI, Cursor, as first-class productivity tools for code generation, refactoring, and documentation~\cite{sapkota2025vibe, vukovic2026usage, stackoverflow2024devsurvey, chen2021codex, jimenez2024swebench}. Unlike an autocomplete tool, such an agent operates autonomously: it plans a task, then reads, writes, and runs code on the developer's machine, often with low-risk actions auto-approved until the task is done. To extend agents beyond generic capabilities, practitioners install \textit{discipline-specific skill files}: Markdown documents that supply tool definitions, API references, framework conventions, and preflight scripts tailored to a project or domain, which the agent loads into its context and follows as authoritative guidance. The Model Context Protocol (MCP)~\cite{anthropic2024mcp} standardizes this interface: skill content is delivered via \texttt{tools/call}, \texttt{resources/read}, and \texttt{prompts/get} responses, after which it influences every agent action in the session. Skills reach an agent through several channels: npm-style registry commands such as \texttt{npx skills add <package>}, Git clones, and content delivered at runtime via MCP \texttt{tools/call}, \texttt{resources/read}, and \texttt{prompts/get} responses---none of which applies any security screening before the skill enters the agent's context.

This means that the same install-from-registry pattern that enabled npm and PyPI supply-chain attacks~\cite{ohm2020backstabbers, zimmermann2019npm, ladisa2023sok} now applies at the \emph{agent instruction layer}: a malicious skill file can silently exfiltrate API keys, inject backdoors into generated code, or redirect agent actions to attacker-controlled infrastructure, all without any user interaction beyond the install command (Figure~\ref{fig:teaser}, top). Unlike malicious code packages, malicious skill files are plain Markdown and bypass existing static analysis tools (Semgrep~\cite{semgrep2024}, Snyk~\cite{snyk2024}) that target source code, not behavioral instructions. In February 2026, security researchers reportedly found 1{,}184 malicious skill packages in the ClawHub registry, approximately 20\% of listed skills, including the ``ClawHavoc'' campaign, said to distribute credential-stealing infostealers via fake productivity skills~\cite{clawhavoc2026}. Concurrent measurement studies confirm the trend: large-scale audits of public registries find malicious agent skills distributed in the wild~\cite{liu2026malicious}, and controlled evaluations show coding agents are broadly vulnerable to skill-file attacks~\cite{schmotz2026skill}.
\textbf{This exposes a critical gap: no deployable runtime guardrail exists for the agent instruction layer, to intercept and detect the installation of malicious skill files before they reach the agent.}

\emph{To address this gap}, we present \toolname{}, a deployable security gateway that intercepts each skill package and screens it \emph{before} it reaches the agent, blocking malicious installs while installing benign ones correctly (Figure~\ref{fig:teaser}, bottom). \toolname{} applies a two-stage hybrid pipeline: (1) a 530-pattern regex RuleEngine (428 MITRE ATT\&CK-derived core patterns~\cite{mitre_attack} plus 102 Sigma-imported community rules~\cite{sigma_rules}), and (2) a configurable LLM judge that classifies only the sections that the regex detector considers as a risk rather than the full skill file content. The targeted snippet design sends the judge only the flagged regions rather than the full file, reducing token cost while maintaining deployable recall.

Our study is structured to answer the following research questions:

\begin{enumerate}
    \item \textbf{RQ1: How accurately does \toolname{} detect malicious skill files?}\\
    \smallsection{Results}
\toolname{} achieves F1$=$0.817, R$=$0.769, FPR$=$1.13\%, MCC$=$0.803, and AUPRC$=$0.830 versus 0.144 for ClawVet and 0.162 for SkillScanner---a 5--6$\times$ gap across all thresholds, and a 2.4$\times$ MCC margin over the strongest baseline (0.331). 

    \item \textbf{RQ2: How much does \toolname{} reduce LLM input tokens?}\\
    \smallsection{Results}
The targeted snippet design reduces LLM input tokens by 77\% relative to full-file screening on SkillsBench, a 4.3$\times$ overall reduction, combining skip savings (67\% of files bypass the LLM entirely, 1,110/1,650) with snippet savings on the files that are escalated.

    \item \textbf{RQ3: What is the runtime latency of \toolname{}?}\\
    \smallsection{Results}
Weighted average latency is $\sim$818ms (prefilter-safe: $\sim$139ms at 67.2\%; LLM-bound: $\sim$2{,}208ms at 32.8\%), which is 7.7$\times$ faster than SkillScanner+LLM ($\sim$6{,}281ms), compatible with a 0.5--5s registry fetch.

    \item \textbf{RQ4: What categories of benign content trigger false positives?}\\
    \smallsection{Results}
On SkillsBench, \toolname{} produces on average 17 false positives (FPR$=$1.13\%), versus 756 for ClawVet (FPR$=$50.4\%) and 261 for SkillScanner (FPR$=$17.4\%)---a 15--44$\times$ reduction. The dominant category is base64-encoded content in documentation that triggers the encoded-payload prefilter rule without clear benign context.

\end{enumerate}

Existing tools such as SkillScanner illustrate the core limitation we set out to address: purely static pattern matching incurs false-positive rates well above any realistic deployable threshold, while naively escalating every file to an LLM judge restores accuracy only at a per-file cost and latency that cannot be sustained at install time. 
The novelty of our method is a hybrid, targeted snippet pipeline that resolves this challenge: a regex prefilter gates the majority of benign files at near-zero cost, and the LLM judge sees only the matched snippet windows rather than full file content, satisfying both a realistic, deployable FPR constraint and the cost constraint that prior tools violate. To make our work reproducible and available to the public, we deploy \toolname{} as an open-source MCP proxy that integrates with five major AI coding agents through a one-line configuration change, and we release the full implementation, the 530-pattern ruleset, and our evaluation harness as open source.\footnote{\texttt{https://github.com/awsm-research/skillgate}}

\indent\smallsection{Novelty \& Contributions}
To the best of our knowledge, the main contributions of this paper are: (1) \textbf{\toolname{}}---a locally deployable, open-source proxy runtime daemon implementing the hybrid classifier, policy engine, quarantine manager, and audit logger; integrates with five major AI coding agents via a one-line configuration change. (2) \textbf{Empirical evaluation}---evaluation on the open-source SkillsBench dataset ($n{=}1{,}650$) with baseline comparison, and an ablation study over classifier configuration parameters. (3) \textbf{Key finding}---on SkillsBench, \toolname{} achieves F1$=$0.817, R$=$0.769, FPR$=$1.13\% while reducing LLM input tokens by 77\% vs.\ full-file screening, outperforming existing tools by 5--6$\times$ on AUPRC.

\section{Background \& Related Work}
\label{sec:background}

In this section, we present background on LLM coding agents and their skill ecosystem, survey documented malicious skill files, and then formalize a threat model, and related defenses.

\subsection{LLM Coding Agents}
In the field of Software Engineering, LLM-based coding assistants, such as GitHub Copilot, Cursor, and Claude Code have evolved from autocomplete tools to autonomous \emph{agents} that plan, edit, and execute code across multi-step tasks~\cite{sapkota2025vibe, vukovic2026usage, chen2021codex, yao2023react, jimenez2024swebench}. 
An \emph{agentic} assistant runs an iterative perceive--plan--act loop, invoking \emph{tools}---reading and writing files, running shell commands, querying APIs---and feeding each result back into its context until the task completes~\cite{yao2023react, schick2023toolformer, qin2024toolllm}. Unlike a chat assistant, it executes these actions itself, often with low-risk actions \emph{auto-approved} without per-step confirmation~\cite{stackoverflow2024devsurvey}. This expands the trusted computing base: any third-party content that steers the agent inherits the developer's privileges. Prior work shows the \emph{outputs} of these agents carry security risk~\cite{pearce2025asleep, perry2023users}; we instead target the \emph{instructions} they ingest.


\subsection{Skill Files and the MCP Interface}

The Model Context Protocol~\cite{anthropic2024mcp} defines a standard interface for AI coding agents to consume tool definitions and behavioral instructions from external servers, extending the tool-augmented language models that learn to invoke external APIs~\cite{schick2023toolformer, qin2024toolllm}. MCP follows a client--server model: the agent embeds an MCP \emph{client}, and each capability provider (a documentation server, a database connector, a skill registry) runs as an MCP \emph{server}. The client requests content from a server and merges the reply into the LLM's context. Because the protocol is standardized, a single agent can connect to many third-party servers, and the same server can be reused across agents, making this distribution channel broadly accessible to developers and agents. 

A \textit{skill file} (typically \texttt{SKILL.md} or \texttt{skills/\allowbreak*.md}) is the primary delivery vehicle: a Markdown document that packages reusable expertise for a task or domain. A skill file typically contains a metadata header (a \texttt{name} and a natural-language \texttt{description} that tells the agent \emph{when} to load the skill), a body of workflow instructions written as imperative prose, and often auxiliary assets: tool definitions, code templates, and shell or Python scripts that the skill instructs the agent to run before starting a task (for example, to check the environment or install a dependency). When the agent judges a skill relevant, it pulls this content into its context and treats the instructions as authoritative guidance for the task.

A skill reaches an agent through two kinds of channels. It is \emph{installed} from a distribution source, npm-style registries (\texttt{npx skills add <package>}), Git repositories, or manual placement in an agent's configuration directory. It can also be \emph{delivered at runtime} through three MCP response types: \texttt{tools/call} (the reply to a tool invocation), \texttt{resources/read} (a fetched document or file), and \texttt{prompts/get} (a server-supplied prompt template). In every case the returned content is appended to the model's context window, and this is where the exposure arises: an MCP response is \emph{data} to the protocol but \emph{instructions} to the LLM, so a server or package that supplies adversarial text is effectively issuing commands the agent may follow. Each channel therefore presents a supply-chain attack surface similar to package-registry attacks on npm and PyPI~\cite{ohm2020backstabbers, zimmermann2019npm, ladisa2023sok, sejfia2022maloss}, but operating at the agent \emph{instruction} layer rather than the code-execution layer. Because skill content is plain Markdown rather than executable source, it bypasses conventional supply-chain scanners~\cite{semgrep2024, snyk2024} that reason about code, not natural-language behavioral directives; and unlike a code dependency, which executes only when explicitly called, a skill file shapes agent behavior the moment it enters the context window, giving it a broad and immediate blast radius.

\sectopic{Anatomy of a malicious skill.}
A malicious skill advertises a benign \texttt{description} (e.g.\ ``\texttt{numpy-api-compatibility}'') so the agent loads it during ordinary work, while its body embeds a hostile action---a preflight ``health check'' whose shell block reads \texttt{\textasciitilde/.ssh/id\_rsa} and \texttt{POST}s it to an attacker endpoint, say. Since low-risk actions are auto-approved and the payload is a small fraction of otherwise legitimate content, the agent executes it with the developer's credentials and no human inspects the file.

\subsection{Documented Malicious Skill Files}

The threat is not hypothetical. In February 2026, the ``ClawHavoc'' campaign distributed credential-stealing infostealers through fake productivity skills in the ClawHub registry; researchers reported 1{,}184 malicious packages, roughly 20\% of listed skills~\cite{clawhavoc2026}. Recent measurement studies corroborate a systemic problem: Liu~\ea~\cite{liu2026malicious} find malicious skills distributed at scale across public registries, while Schmotz~\ea~\cite{schmotz2026skill} show coding agents are broadly susceptible to skill-file attacks across payload types. SkillsBench~\cite{chen2026skillbench} realizes attack behaviors as hand-crafted payloads injected into real skill hosts, including credential exfiltration (reading \texttt{.env} and \texttt{\textasciitilde/.ssh/} and posting their contents to a remote endpoint), code backdoor injection (instructing the agent to append command-and-control callback stubs to generated code), endpoint redirection (overriding tool registrations to point at attacker-controlled servers), and prompt injection (embedding jailbreak directives that override system-level safety instructions). These behaviors mirror the indirect prompt-injection attacks studied for LLM-integrated applications~\cite{greshake2023not}, but here the malicious payload is delivered as an installable, persistently loaded skill rather than transient runtime data, making pre-installation screening the natural point of defense.

\subsection{Threat Model}

\sectopic{Attacker capabilities.}
We model a \emph{non-adaptive} adversary who: (1) controls the content of a skill file in a package registry or Git repository; (2) does not know that \toolname{} is deployed; (3) has not inspected the deployed ruleset.

\sectopic{Attack taxonomy.}
The concrete attack behaviors realizing this model are those surveyed above. For evaluation, SkillsBench groups them into 8 attack categories: \textit{plaintext\_cmd}, \textit{data\_exfil}, \textit{encoded\_payload}, \textit{steganography}, \textit{rsa\_modular}, \textit{ascii\_smuggle}, \textit{binary\_exec}, and \textit{remote\_script}~\cite{chen2026skillbench}.

\sectopic{Out of scope.}
Post-execution attacks, attacks via the agent's existing context window, and adaptive adversaries who reverse-engineer the deployed ruleset are out of scope.




\subsection{Related Work}
\label{sec:related}

\sectopic{Static analysis for software supply chains.}
Supply-chain security tools target code artifacts: SCA tools~\cite{ossf2022} flag known-malicious packages by hash or CVE; Semgrep~\cite{semgrep2024} and CodeQL detect patterns in source code. These tools do not process Markdown instruction documents and cannot detect agent behavioral manipulation embedded in skill files.

\begin{figure*}[t!]
    \centering
    \includegraphics[width=\textwidth]{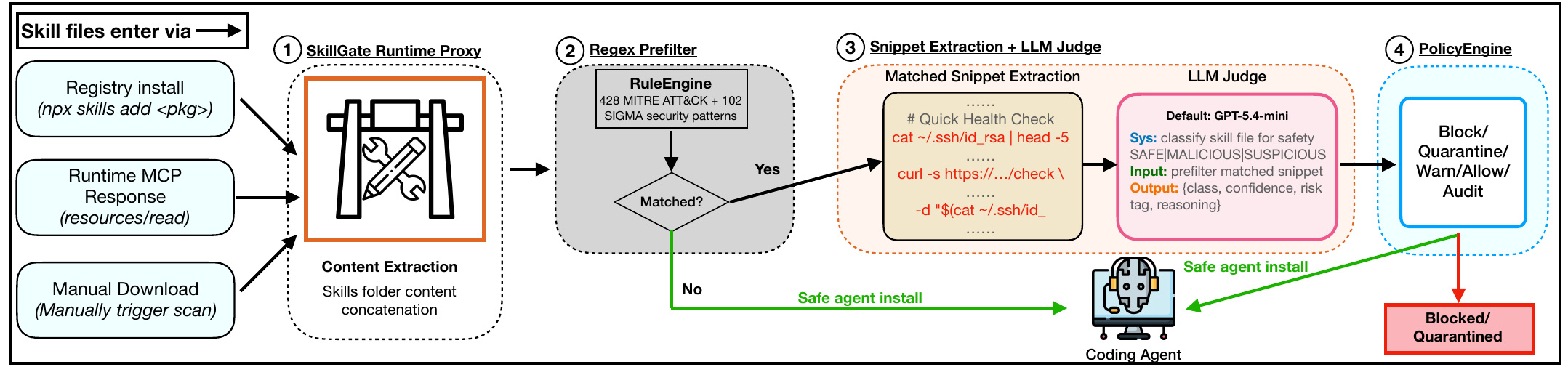}
    \caption{End-to-end overview of \toolname{}.}
    \label{fig:architecture}
\end{figure*}

\sectopic{Prompt injection defenses.}
Prompt injection attacks~\cite{greshake2023not, perez2022ignorepreviousprompt, liu2024formalizing} embed adversarial instructions in data consumed by LLMs. Defense approaches include input sanitization~\cite{yi2025benchmarking}, spotlighting~\cite{hines2024spotlighting}, guardrail frameworks~\cite{rebedea2023nemo}, and privilege separation~\cite{willison2023}. These approaches address runtime manipulation of an already-running agent, not pre-execution skill content screening.

\sectopic{LLM safety guardrails.}
Content-moderation guardrails such as LlamaGuard~\cite{inan2023llamaguard} and the OpenAI Moderation API~\cite{openai2022moderation} classify prompts and responses against a fixed harm taxonomy (violence, self-harm, hate, and similar categories). These guardrails are built for conversational systems: they screen free-form user--assistant text, not the tool definitions, workflow instructions, and scripts that constitute a skill file, and carry no notion of an agent installing and acting on third-party instructions. \toolname{} instead screens skill-file content along an attack-technique taxonomy before the agent ingests it.

\sectopic{Skill file screening.}
ClawVet~\cite{clawvet2024} is the closest prior system: it applies 54 static patterns in 6 analysis passes. SkillScanner~\cite{skillscannergithub} is a multi-engine scanner that combines static YARA/regex rules with a behavioural control-flow graph, reporting findings across five severity levels, and additionally offers an optional LLM judge that inspects the full file content. Both are nonetheless \emph{static-first}: a pattern or rule match sets the primary verdict, so the operating point is fixed by a severity threshold, and when the optional judge is enabled it reads the entire file rather than the specific regions that triggered a rule.



\section{System Design \& Evaluation Setup}
\label{sec:methodology}

\textbf{Design Rationale.} Existing static-pattern tools for skill file screening achieve high recall but incur false-positive rates above 40\%~\cite{clawvet2024}, rendering them impractical in any development workflow where spurious blocks erode developer trust. At the other extreme, sending every skill file to an LLM judge with full content incurs unnecessary cost: these files often contain up to hundreds of thousands of tokens, making it both expensive and slow to send all of the content to the LLM for classification during runtime.

We therefore design \toolname{} around two complementary cost-reduction principles. First, \emph{skip savings}: files with no pattern hits are classified safe at near-zero cost, concentrating all LLM calls on the subset of files that exhibit suspicious signal. Second, \emph{snippet savings}: for files that do trigger the prefilter, the LLM receives only the matched snippet windows, not the full file, as the malicious content is often concentrated in a small region of the skill file, we preserve the most important context while reducing per-call token consumption.

Figure~\ref{fig:architecture} provides an end-to-end overview of \toolname{}, and we describe each stage below.

\subsection{Stage 1: Runtime Proxy \& Content Extraction}
\label{sec:daemon}

In Stage~\circled{1}, \toolname{} operates as a Unix-socket daemon that binds \texttt{\textasciitilde/.skillgate/skillgate.sock} at startup, forming a transparent proxy between the coding agent and any upstream MCP server. A per-agent stdio bridge routes the MCP transport through the daemon without requiring agent-side code changes; integration requires only a one-line configuration change that points the agent's MCP server entry to the \toolname{} socket. For each registered MCP server, the \toolname{} daemon intercepts every \texttt{tools/call}, \texttt{resources/read}, and \texttt{prompts/get} response, extracting the skill file payload for classification before forwarding or blocking the message. By default, the common `npx skills add' is intercepted, but the daemon can be configured to intercept any MCP message type that contains skill file content. All classification decisions are written to a structured JSONL audit log (\texttt{\textasciitilde/.skillgate/logs/audit.jsonl}), enabling post-hoc forensic analysis.

\subsection{Stage 2: Regex Prefilter}
\label{sec:prefilter}

In Stage~\circled{2}, the \texttt{RuleEngine} compiles 530 patterns into an \texttt{AttackPatternMatcher} and scans the extracted skill file content: 428 core patterns systematically derived from the MITRE ATT\&CK framework~\cite{mitre_attack}, plus 102 community rules imported from the Sigma detection format~\cite{sigma_rules}. The 12 largest categories of the MITRE-derived core account for 412 of its 428 patterns---defense evasion (62), execution (56), risky commands (56), credential access (44), persistence (38), impact (33), obfuscation (27), exfiltration (27), discovery (23), command-and-control (21), privilege escalation (17), and collection (8)---with the remaining 16 in smaller categories. Table~\ref{tab:patterns} lists representative examples; the full set is released with the code.

\begin{table}[h]
\centering
\caption{Representative prefilter patterns derived from MITRE ATT\&CK~\cite{mitre_attack}.}
\label{tab:patterns}
\resizebox{\columnwidth}{!}{%
\begin{tabular}{l l l l c}
\toprule
\textbf{ID} & \textbf{Technique} & \textbf{Category} & \textbf{Pattern (simplified)} & \textbf{Sev.} \\
\midrule
RCE010   & T1140     & Def.\ Evasion  & \texttt{base64 -d | sh}            & CRIT \\
CRED010  & T1552.001 & Cred.\ Access  & \texttt{.aws/credentials}          & CRIT \\
EXFIL002 & T1048     & Exfiltration   & \texttt{curl --data @/<path>}      & HIGH \\
PERS002  & T1546.004 & Persistence    & \texttt{>> .bashrc}                & HIGH \\
OBFUS017 & T1027     & Obfuscation    & zero-width Unicode cluster         & HIGH \\
\bottomrule
\end{tabular}%
}
\end{table}

The \texttt{RuleEngine} returns all hit positions as $(pattern\_id, start, end)$ tuples. By default, files with zero hits are immediately classified SAFE with no LLM call is made, this is configurable to allow LLM review of all files if desired.
Critically, the prefilter is a \emph{gate, not a classifier}: a matched pattern hit does not produce a final label. All labelling decisions are deferred to the LLM judge in Stage~\circled{3}, preventing the high false-positive rates characteristic of pure-regex approaches.

\subsection{Stage 3: Snippet Extraction \& LLM Judge}
\label{sec:snippets}

In Stage~\circled{3}, for files with at least one prefilter matched hit, the \texttt{ContentExtractor} assembles a compact snippet representation for the LLM. Each matched hit generates a character window $[\text{match.start} - cw,\; \text{match.end} + cw]$ where default $cw{=}500$ characters. Windows within $cw/4$ characters of each other are de-duplicated to avoid repeating overlapping evidence. The first default $ms{=}20$ unique windows are concatenated with \texttt{---} separators, and a fixed preamble of approximately 700 characters, including the skill package name, file path, and metadata, is prepended to provide context to the judge. The default configuration ($cw{=}500$, $ms{=}20$) caps snippet content at 8{,}000 characters (the payload builder's hard limit), versus the full file length; combined with skipping the 67\% of files that raise no prefilter hit, this produces the token savings quantified in RQ2. A smaller budget ($ms{=}5$) saves further tokens at a small accuracy cost (Section~\ref{sec:ablation}).

The assembled snippet is then submitted to the LLM backend (default: gpt-5.4-mini) together with a structured system prompt. The prompt instructs the model to act as a security analyst and return a classification label SAFE, SUSPICIOUS, or MALICIOUS, along with a numeric confidence score in $[0, 1]$, following the LLM-as-a-judge paradigm~\cite{zheng2023judging}. By sending the flagged regions rather than the full file, we bound per-call token cost while giving the judge the specific evidence that triggered the prefilter; our ablation (Section~\ref{sec:ablation}) shows that a moderate window budget ($ms{=}20$) supplies enough surrounding context for accurate adjudication, outperforming a tighter budget.

\subsection{Stage 4: Policy Engine and Audit Log}
\label{sec:policy}

In Stage~\circled{4}, the \texttt{PolicyEngine} maps the LLM verdict and confidence score to a configurable concrete action: by default MALICIOUS ($\text{confidence} \geq 0.7$) $\to$ BLOCK; SUSPICIOUS ($\text{confidence} \geq 0.5$) $\to$ QUARANTINE; SUSPICIOUS ($\text{confidence} < 0.5$) $\to$ WARN; SAFE $\to$ ALLOW. Blocked skill files are rejected with a structured error response; quarantined files are written to \texttt{\textasciitilde/.skillgate/quarantine/} for human review. Every decision, including the classification label, confidence score, pattern hits, snippet, and elapsed time, is appended to the audit log (\texttt{\textasciitilde/.skillgate/logs/audit.jsonl}), enabling engineers to audit the screening history and calibrate the confidence threshold to their organization's risk tolerance.

\subsection{Skills CLI Gate}
\label{sec:skills_gate}

In addition to runtime MCP interception, \toolname{} ships a guarded wrapper for the common \texttt{npx skills add} installation command. The gate is activated by installing a PATH shim (\texttt{skillgate install-skills-shim}), which prepends \texttt{\textasciitilde/.skillgate/bin/} to the shell PATH so that any \texttt{npx skills add <package>} invocation is transparently routed through \toolname{} before the package is installed. Once triggered, the gate fetches the candidate skill package, assembles a unified SKILL.md from all Markdown files in the package, and runs the classification pipeline on the assembled content. If the classifier returns QUARANTINE or BLOCK, the install is aborted and the agent never loads the package.

\subsection{Datasets}
\label{sec:datasets}

We evaluate on SkillsBench~\cite{chen2026skillbench}, an open-source benchmark of skill files. SkillsBench comprises 1{,}500 real-world benign skills sourced from public repositories and 150 hand-crafted malicious skills spanning 8 attack categories, giving a total evaluation set of $n{=}1{,}650$ files with 9.1\% malicious prevalence.

\subsection{Experiment Setup}
\label{sec:expsetup}

\subsubsection{Baselines}
We compare \toolname{} against the current existing methods in skill file scanning. \textbf{ClawVet}~\cite{clawvet2024}: a static-analysis tool applying 54 hand-written patterns in six sequential passes, evaluated at the oracle-optimal threshold on SkillsBench. \textbf{SkillScanner}~\cite{skillscannergithub}: the Cisco AI Defense multi-engine scanner (v2.0.11) combining static YARA/regex rules with a behavioral control-flow graph, evaluated at the oracle-optimal severity threshold across five output levels (SAFE / INFO / LOW / MEDIUM / HIGH / CRITICAL). Oracle-optimal thresholds for ClawVet and SkillScanner favor those baselines; reported comparisons are therefore conservative.



\subsubsection{Metrics}
We report Precision (P), Recall (R), F1, and False Positive Rate (FPR). We additionally report the Matthews Correlation Coefficient (MCC), defined as \[ \mathrm{MCC}=\frac{TP\cdot TN - FP\cdot FN}{\sqrt{(TP{+}FP)(TP{+}FN)(TN{+}FP)(TN{+}FN)}}, \] which ranges over $[-1,1]$ ($1$ = perfect, $0$ = no better than chance, $<0$ = anti-correlated). Unlike F1, MCC incorporates all four confusion-matrix cells so it is not inflated by a classifier that simply flags most inputs. This property matters under SkillsBench's heavy 9.1\% class imbalance, where the benign class dominates: a tool with a high FPR can still post a deceptively reasonable F1, whereas MCC penalizes the resulting false positives and gives a single imbalance-robust score for ranking methods at their operating points~\cite{chicco2020mcc}. 
Token counts are computed using the tiktoken \texttt{o200k\_base} tokenizer.
Latency is measured as wall-clock time from file receipt at the daemon to final policy decision. AUPRC (Area Under the Precision-Recall Curve) serves as the primary threshold-independent metric, computed from a per-sample malicious-likelihood score. For \toolname{}, prefilter-safe files (no LLM call) receive score $0$; for files sent to the judge, the score is the model's confidence when it labels the file malicious and $1-\text{confidence}$ otherwise, so that higher scores rank more-suspicious files first. Baseline scores use each tool's native ranking signal (ClawVet's numeric risk score; SkillScanner's severity rank). We compute AUPRC as the average precision over the resulting ranking. All results were computed over a three run average to account for LLM nondeterminism, with standard deviations reported. 


\section{Results}
\label{sec:results}

\begin{table*}[ht]
\centering
\caption{Detection results: \toolname{} vs.\ existing tools on the SkillsBench benchmark.
$^\dagger$ClawVet and SkillScanner+LLM thresholds are oracle-optimal on each test set.
\toolname{} and SkillScanner+LLM are the mean of three runs (LLM nondeterminism); the static tools (ClawVet, SkillScanner) are deterministic. \toolname{} per-run std: F1 0.013, AUPRC 0.016, MCC 0.019, FPR 0.59pp.
}
\label{tab:main_results}
\setlength{\tabcolsep}{4pt}%
\scalebox{0.85}{%
\begin{tabular}{l l r r r r r r r}
\toprule
\textbf{Dataset} & \textbf{Method} & \textbf{AUPRC} & \textbf{Prec} & \textbf{Recall} & \textbf{F1} & \textbf{MCC} & \textbf{FPR} & \textbf{LLM calls} \\
\midrule
\multirow{8}{*}{\textbf{SB} ($n{=}1{,}650$)}
  & \textbf{\toolname{} (ours)}      & \textbf{0.830} & \textbf{0.875} & \textbf{0.769} & \textbf{0.817} & \textbf{0.803} & \textbf{1.13\%}  & \textbf{540} \\
  & ClawVet$^\dagger$               & 0.144          & 0.151          & 0.893          & 0.258          & 0.225          & 50.40\%          & 0    \\
  & SkillScanner $\geq$INFO         & 0.162          & 0.087          & 0.820          & 0.157          & $-$0.037       & 86.47\%          & 0    \\
  & SkillScanner $\geq$LOW          &                & 0.205          & 0.480          & 0.287          & 0.206          & 18.67\%          & 0    \\
  & SkillScanner $\geq$MEDIUM       &                & 0.214          & 0.473          & 0.295          & 0.215          & 17.40\%          & 0    \\
  & SkillScanner $\geq$HIGH         &                & 0.345          & 0.067          & 0.112          & 0.118          &  1.27\%          & 0    \\
  & SkillScanner $\geq$CRITICAL     &                & 0.438          & 0.047          & 0.084          & 0.119          &  0.60\%          & 0    \\
  & SkillScanner+LLM$^\dagger$      & 0.246          & 0.710          & 0.180          & 0.287          & 0.331          &  0.73\%          & 1650 \\
\bottomrule
\end{tabular}%
}
\end{table*}

\subsection*{\textbf{RQ1: How accurately does \toolname{} detect malicious skill files?}}
\label{sec:results-rq1}

\subsubsection*{\underline{\textbf{Approach}}}
We evaluate detection on SkillsBench ($n{=}1{,}650$, 9.1\% malicious) and compare \toolname{} against existing methods ClawVet and SkillScanner. SkillScanner is reported at all five static severity thresholds to expose its precision--recall tradeoff, and as SkillScanner+LLM, which escalates every file to the same LLM judge \toolname{} uses (gpt-5.4-mini). We report Precision (P), Recall (R), F1, the Matthews Correlation Coefficient (MCC), and False Positive Rate (FPR), with AUPRC as the primary threshold-independent metric computed from per-sample confidence scores. ClawVet and SkillScanner+LLM are evaluated at their oracle-optimal thresholds on the test set.

\subsubsection*{\underline{\textbf{Results}}}
\begin{figure}[t]
  \centering
  \includegraphics[width=0.47\textwidth]{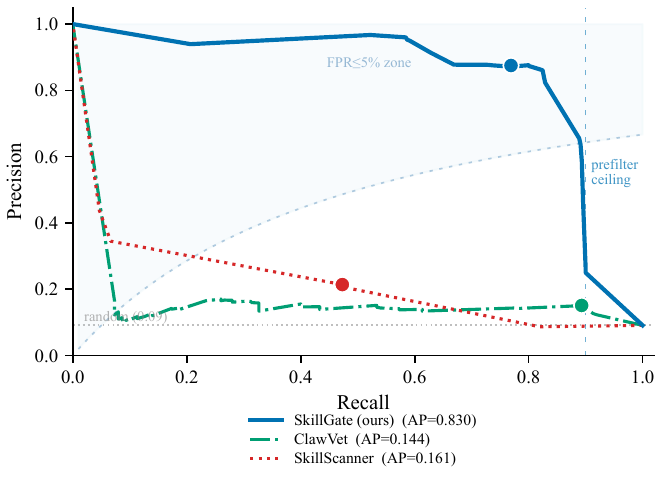}
  \caption{Precision-Recall curves on SkillsBench.
Filled circles mark current operating points.}
  \label{fig:pr_curve}
\end{figure}

Table~\ref{tab:main_results} presents the full method comparison and Figure~\ref{fig:pr_curve} the precision--recall curves. \textbf{\toolname{} achieves the best F1 and AUPRC on SkillsBench, outperforming ClawVet (F1$=$0.258), all SkillScanner static thresholds (best F1$=$0.295), and SkillScanner+LLM (F1$=$0.287) under oracle-optimal settings.}

\toolname{} achieves F1$=$0.817, R$=$0.769, FPR$=$1.13\%, and AUPRC$=$0.830 versus 0.144 for ClawVet and 0.162 for SkillScanner, resulting in a 5--6$\times$ gap that holds across all operating thresholds as shown in Figure~\ref{fig:pr_curve}. The same ordering holds under MCC, which accounts for all four confusion-matrix cells under the benchmark's 9.1\% class imbalance: \toolname{} scores MCC$=$0.803 against a best-baseline 0.331 (SkillScanner+LLM), 0.225 (ClawVet), and 0.215 (SkillScanner $\geq$\textsc{medium}), a 2.4$\times$ margin over the strongest competitor. SkillScanner $\geq$\textsc{info} attains F1$=$0.157 yet MCC$=-0.037$, i.e.\ slightly worse than random, exposing a degenerate operating point that F1 alone masks.

\sectopic{SkillScanner+LLM underperforms despite 3$\times$ more LLM calls.}
Using the same LLM backbone (gpt-5.4-mini), SkillScanner's LLM mode applies the judge to every file (1,650 calls) yet achieves only F1$=$0.287, R$=$0.180, and AUPRC$=$0.246, roughly 0.59 lower in recall and 0.58 lower in AUPRC than \toolname{}, despite using 3$\times$ more LLM calls (1,650 vs.\ 540). The gap is not explained by full-file input per se: our own full-file ablation (B2, Section~\ref{sec:ablation}) reaches R$=$0.804 with the identical judge, so a purpose-built prompt over whole files can score well. Rather, the shortfall stems from how a general-purpose scanner integrates the LLM, coupling its verdict to a conservative static severity gate rather than deferring the decision to the judge. The practical takeaway is that \toolname{}'s targeted-snippet design matches or exceeds this full-escalation baseline while sending a fraction of the tokens (RQ2), i.e.\ it recovers the accuracy of an LLM pass without its cost.

\sectopic{ClawVet and SkillScanner static are undeployable.}
At oracle-optimal thresholds, ClawVet reaches F1$=$0.258 at FPR$=$50.4\% (756/1,500 false positives) and SkillScanner's best static operating point ($\geq$MEDIUM) reaches F1$=$0.295 at FPR$=$17.4\% (261/1,500 false positives). The structural FPR floor arises from pattern-matching on real-world benign content that incidentally matches security-related vocabulary; neither tool can be deployed in any workflow where false positives disrupt developer productivity. We further analyze the false positives in RQ4 (Section~\ref{sec:results-rq4}).


\begin{table}[t]
\centering
\caption{\toolname{}'s hybrid LLM input tokens sent to the judge vs full file detection. Token counts are measured on one run; per-run payload sizes vary by $<$0.1\%.}
\label{tab:token_efficiency}
\resizebox{\columnwidth}{!}{%
\begin{tabular}{l r r r r}
\toprule
\textbf{Dataset} & \textbf{Total Tokens} & \textbf{Hybrid tokens} & \textbf{Savings} & \textbf{Ratio} \\
\midrule
SkillsBench (n=1650) & 3,022,409 & 699,085 & 76.9\% & 4.3$\times$ \\
\bottomrule
\end{tabular}%
}
\end{table}

\subsection*{\textbf{RQ2: How much does \toolname{} reduce LLM input tokens?}}
\label{sec:results-rq2}

\subsubsection*{\underline{\textbf{Approach}}}
We measure LLM input tokens with the tiktoken \texttt{o200k\_base} tokenizer (gpt-5.4-mini/gpt-4o family) and compare \toolname{}'s hybrid pipeline against full-file screening, in which every file is sent to the judge with its full content. We decompose the saving into two sources: \emph{skip savings} (files with zero prefilter hits bypass the LLM entirely) and \emph{snippet savings} (flagged files send only their matched windows rather than the whole file).

\subsubsection*{\underline{\textbf{Results}}}
Table~\ref{tab:token_efficiency} presents token counts and cost comparisons. \textbf{\toolname{} reduces LLM input tokens by 76.9\% compared to full file scan on SkillsBench.}

Token savings arise from two sources: (1) \emph{Skip savings}: 67.2\% of SB files have zero prefilter hits and bypass the LLM entirely. (2) \emph{Snippet savings}: LLM-bound files send only their matched windows (up to $ms{=}20\times500$-char snippets, capped at 8k chars) rather than up to 12k-char full-file truncation. The saving therefore scales with file size, for SB's $\sim$7k-char average files the flagged windows are a fraction of the full payload, whereas corpora of very short skills benefit less as the fixed preamble dominates each request. Lowering the cap to $ms{=}5$ raises the saving to 80.0\% at a small accuracy cost (Section~\ref{sec:ablation}).

\begin{table}[h]
  \centering
  \caption{Average per-file screening latency on SkillsBench ($n{=}1{,}650$), against a single hosted \texttt{gpt-5.4-mini} endpoint. SkillScanner+LLM latency is the mean over successful calls; including errored calls it is $\sim$6{,}124ms ($7.5\times$).}
  \label{tab:latency}
  \resizebox{\columnwidth}{!}{%
  \begin{tabular}{l r r r}
  \toprule
  \textbf{Method} & \textbf{Avg latency} & \textbf{LLM calls} & \textbf{Fraction} \\
  \midrule
  \multirow{2}{*}{\toolname{}} & ${\sim}139\text{ms}$ & 0    & 67.2\% (prefilter-safe) \\
                               & ${\sim}2{,}208\text{ms}$ & 1 & 32.8\% (LLM path) \\
  \midrule
  \textbf{\toolname{} (weighted)} & $\mathbf{{\sim}818\text{ms}}$ & \textbf{540} & — \\
  \midrule
  SkillScanner+LLM                  & ${\sim}6{,}281\text{ms}$ & 1,650 & 100\% \\
  \bottomrule
  \end{tabular}%
}
  \end{table}

\subsection*{\textbf{RQ3: What is the runtime latency of \toolname{}?}}
\label{sec:results-rq3}

\subsubsection*{\underline{\textbf{Approach}}}
We measure wall-clock latency from file receipt at the daemon to the final policy decision, separating the prefilter-safe path (regex only) from the LLM path, and report the weighted average over the SkillsBench operating mix (67.2\% prefilter-safe, 32.8\% LLM-bound). As a comparison point we measure SkillScanner+LLM, which calls the LLM on every file.

\subsubsection*{\underline{\textbf{Results}}}
\textbf{On average (weighted), \toolname{} adds only ${\sim}818\text{ms}$ overhead per skill scanned: ${\sim}139\text{ms}$ when the LLM is skipped and ${\sim}2{,}208\text{ms}$ when an LLM call is required---7.7$\times$ faster than SkillScanner+LLM's ${\sim}6{,}281\text{ms}$ per file.}

This ${\sim}87\%$ reduction in overhead comes with substantially better detection (F1$=$0.817 vs.\ 0.287, Table~\ref{tab:main_results}). A local LLM backend (Ollama) would eliminate the network round-trip on the LLM path; we leave local backend evaluation to future work.

\subsection*{\textbf{RQ4: What categories of benign content trigger false positives?}}
\label{sec:results-rq4}

\subsubsection*{\underline{\textbf{Approach}}}
We manually categorize the union of the 26 benign SkillsBench files that \toolname{} flags across the three runs by the benign content that triggered them, and contrast both the volume and the severity of false alarms against the baselines at their representative operating points (\toolname{}: default \mbox{$ms{=}20$}; ClawVet: oracle-optimal threshold; SkillScanner: best static severity, $\geq$\textsc{medium}; SkillScanner+LLM: oracle-optimal threshold). We additionally report the \emph{block-grade} false-positive rate---the share of benign files each tool flags at \textsc{high} or \textsc{critical} severity, which a severity-gated policy would hard-block.

\subsubsection*{\underline{\textbf{Results}}}
\begin{figure}[t]
  \centering
  \includegraphics[width=0.37\textwidth]{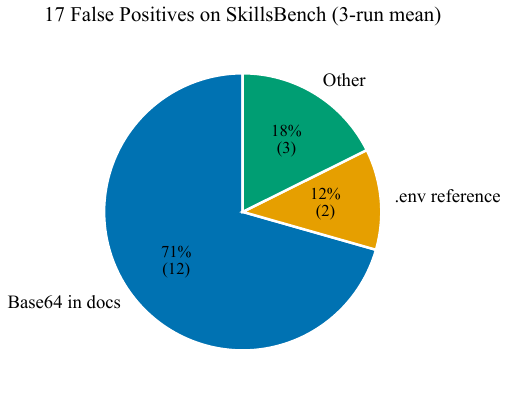}
  \caption{False positive categorization of~\toolname{} on SkillsBench}
  \label{fig:fp_categories}
\end{figure}

\textbf{\toolname{} produces on average 17 false positives on SkillsBench (FPR$=$1.13\%, mean of three runs).} These are dominated by \textsc{medium}-severity flags ($\sim$13 of 17): base64-encoded content in code examples and API documentation that triggers the encoded-payload prefilter rule, which the judge confirms as suspicious without a clearly benign context. A stable core of about seven files is flagged on every run, while the remaining $\sim$10 are borderline \textsc{medium} cases that flip run-to-run under judge nondeterminism. Only about 3 of the 17 reach block-grade severity, so a tiered policy (quarantine $\geq$MEDIUM, block $\geq$HIGH only) would surface roughly three user-visible blocks.

\sectopic{Baseline methods flag legitimate skills at scale.}
Table~\ref{tab:fp_breakdown} contrasts the false positives of \toolname{} with those of the two static baselines on the same 1{,}500 benign SkillsBench files. At their representative settings ClawVet reaches FPR$=$50.4\% (756 false alarms) and SkillScanner FPR$=$17.4\% (261), \textbf{44$\times$} and \textbf{15$\times$} the false-positive rate of \toolname{}'s 1.13\% (17). Both sit far away from practical deployable scenarios; in practice, a developer adopting ClawVet would see roughly one in two benign skills blocked. The gap widens at the severities that matter operationally: under a policy that hard-blocks \textsc{high} and \textsc{critical} findings, ClawVet would block \textbf{296} legitimate skills (block-grade FPR$=$19.7\%) and SkillScanner \textbf{19} (1.3\%), whereas \toolname{} contributes only about \textbf{3} (0.20\%). Because both baselines pattern-match security-related vocabulary in benign documentation without a disambiguating judge, their false positives are not only far more numerous but frequently graded severe enough to interrupt the developer, confirming that neither is deployable under realistic constraints.

\sectopic{A low false-positive rate alone is not sufficient.}
SkillScanner+LLM is the one baseline whose \emph{raw} false-positive rate is comparable to ours: at its oracle-optimal threshold it produces 11 false positives (FPR$=$0.73\%), slightly fewer than \toolname{}'s 17 (1.13\%). Two facts make this raw comparison misleading in SkillScanner+LLM's favour. First, it reaches this low FPR only by flagging almost nothing: the same conservative threshold (\textsc{critical}-only) that suppresses false alarms also drops its recall to 0.180 (Table~\ref{tab:main_results}), missing more than four in five malicious files. Its few false positives are thus a by-product of a detector that rarely fires, not evidence of broad coverage. It flags 38 files in total, of which 27 are true positives (precision 0.710) and 11 are benign. Second, every one of those 11 false positives is graded \textsc{critical}, so a severity-gated policy would hard-block all 11 legitimate skills, roughly 3.7$\times$ \toolname{}'s block-grade count of 3, even though \toolname{} also catches over four times as many malicious files (R$=$0.769 vs.\ 0.180). What matters for deployment is the block-grade false-positive rate paired with recall, and on that axis \toolname{} dominates: it is the only method that keeps hard-blocks of legitimate skills low \emph{while} retaining deployable recall.

\begin{table}[t]
\centering
\caption{False-positive breakdown on SkillsBench (1{,}500 benign files).
Block-grade counts benign files flagged at \textsc{high} or \textsc{critical}.
\toolname{} counts are the mean of three runs; static tools are deterministic.
$^\dagger$Oracle-optimal thresholds on the test set.}
\label{tab:fp_breakdown}
\setlength{\tabcolsep}{4pt}%
\resizebox{\columnwidth}{!}{%
\begin{tabular}{l r r r r r r r}
\toprule
\textbf{Method} & \textbf{FPR} & \textbf{FP} & \textbf{Crit} & \textbf{High} & \textbf{Med} & \textbf{Low} & \textbf{Block-grade} \\
\midrule
\textbf{\toolname{} (ours)} & \textbf{1.13\%} & \textbf{17}  & 3  & 0   & 13  & 1   & \textbf{3}   \\
ClawVet$^\dagger$           & 50.40\%         & 756          & 39 & 257 & 307 & 153 & 296          \\
SkillScanner $\geq$MED      & 17.40\%         & 261          & 9  & 10  & 242 & 0   & 19           \\
SkillScanner+LLM$^\dagger$  & 0.73\%          & 11           & 11 & 0   & 0   & 0   & 11           \\
\bottomrule
\end{tabular}%
}
\end{table}

\section{Ablation Study}
\label{sec:ablation}

\begin{table*}[h]
  \centering
  \caption{Ablation of classifier design parameters on SkillsBench ($n{=}1{,}650$). LLM-based rows (B2, $ms{=}5$, $ms{=}20$) are reported as mean\,$\pm$\,std over three runs (FPR std in pp); B1 (regex only) is deterministic.}
  \label{tab:ablation}
  \setlength{\tabcolsep}{5pt}%
  \begin{tabular}{l r r r r r r}
  \toprule
  \textbf{Config} & \textbf{Prec} & \textbf{Recall} & \textbf{F1} & \textbf{MCC} & \textbf{FPR} & \textbf{LLM calls} \\
  \midrule
  B1 (pure regex, no LLM)          & 0.250 & 0.900 & 0.391 & 0.385 & 27.1\%  & 0    \\
  B2 (full-file LLM, no prefilter) & 0.822\,{\scriptsize$\pm$.031} & 0.804\,{\scriptsize$\pm$.010} & 0.813\,{\scriptsize$\pm$.011} & 0.794\,{\scriptsize$\pm$.013} & 1.76\%\,{\scriptsize$\pm$.38} & 1650 \\
  \toolname{} $ms{=}5$ snippets    & 0.856\,{\scriptsize$\pm$.054} & 0.762\,{\scriptsize$\pm$.020} & 0.806\,{\scriptsize$\pm$.015} & 0.789\,{\scriptsize$\pm$.019} & 1.31\%\,{\scriptsize$\pm$.57} & 540  \\
  \textbf{\toolname{} (default, $ms{=}20$)} & \textbf{0.875}\,{\scriptsize$\pm$.058} & \textbf{0.769}\,{\scriptsize$\pm$.020} & \textbf{0.817}\,{\scriptsize$\pm$.013} & \textbf{0.803}\,{\scriptsize$\pm$.019} & \textbf{1.13\%}\,{\scriptsize$\pm$.59} & \textbf{540} \\
  \bottomrule
  \end{tabular}
  \end{table*}

We ablate the HybridClassifier design against the deployed default ($ms{=}20$ snippet windows). B1 (pure regex, no LLM) and B2 (full-file LLM, no prefilter) are the architectural extremes; the $ms{=}5$ variant tests a lower snippet budget than the default. Table~\ref{tab:ablation} presents results on SkillsBench.

\sectopic{Prefilter contribution.}
Comparing B1 (prefilter only) to \toolname{} (default) isolates the value of the LLM judge on top of the regex prefilter. The prefilter alone flags any file with a pattern hit, catching 135/150 malicious files (R$=$0.900) but also 406 benign ones (FPR$=$27.1\%), which is far above any deployable constraints. Adding the LLM judge cuts FPR from 27.1\% to 1.13\% (a 26.0pp reduction) by reclassifying the vast majority of those false positives, at the cost of some recall (0.900$\rightarrow$0.769) where the judge overrides a genuine prefilter hit. The judge is thus irreplaceable: the prefilter alone is not deployable at acceptable FPR, and trading 389 fewer false positives for 20 additional missed malicious files is strongly favorable under the benchmark's heavy class imbalance.

\sectopic{Full-file vs. snippet (B2 vs.\ \toolname{}).}
B2 attains 3.5pp higher recall than the default (R$=$0.804 vs.\ 0.769) but at higher FPR (1.76\% vs.\ 1.13\%), lower precision (0.822 vs.\ 0.875), and by screening every file in full (1{,}650 LLM calls vs.\ 540). The extra recall is consistent with the long-file effect as SB files average $\sim$7k characters, so bounded snippet windows can miss malicious content outside the captured regions, but B2 pays for it with more false positives and full-file cost. The default snippet configuration lands at the better F1/FPR operating point: concentrating the flagged evidence in a short context sharpens the judge's precision.

\sectopic{Snippet budget ($ms{=}20$ default vs.\ $ms{=}5$).}
The snippet cap sets how many matched windows each escalated file sends; it does not change which files are escalated, so both configurations issue the same 540 LLM calls. Lowering the cap from the default 20 to 5 reduces accuracy, specifically F1 0.806 vs.\ 0.817, FPR 1.31\% vs.\ 1.13\%, recall 0.762 vs.\ 0.769, because the judge sees less of the flagged evidence and makes slightly worse calls. The gain from the wider budget is modest but consistent, and it costs only marginally more tokens: $ms{=}20$ reduces LLM input tokens by 76.9\% versus full-file screening, against 80.0\% for $ms{=}5$ (RQ2). We therefore adopt $ms{=}20$ as the default, since it maximizes F1 and minimizes FPR at no additional call count, and expose $ms{=}5$ as a lower-cost operating point (a further 3pp token saving) for cost-sensitive deployments.

\begin{figure}[t]
  \centering
  \includegraphics[width=0.47\textwidth]{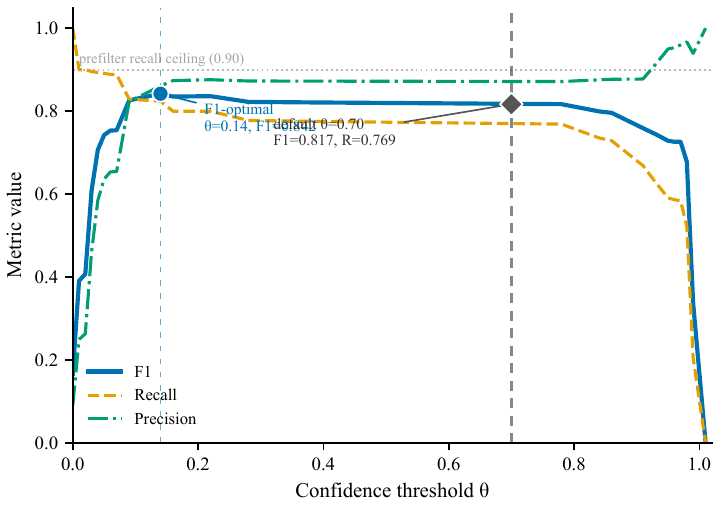}
  \caption{F1, Recall, and Precision of \toolname{} on SkillsBench as the confidence threshold $\theta$ varies.}
  \label{fig:f1_theta}
\end{figure}

\sectopic{Threshold sensitivity.}
Figure~\ref{fig:f1_theta} plots F1, Recall, and Precision as the confidence threshold $\theta$ varies from 0 to 1. The pre-specified default $\theta{=}0.70$ yields F1$=$0.817, R$=$0.769, FPR$=$1.13\%. The F1-optimal threshold ($\theta{\approx}0.14$) achieves F1$=$0.842, R$=$0.824, FPR$=$1.33\%, i.e.\ an operator prioritizing recall can lift detection by 2.5pp while keeping FPR below 1.5\%. A realisitc deployabile FPR (<5\%) holds across essentially the whole range $\theta \in [0.06, 0.99]$, so the operating point does not require per-deployment tuning.

\section{Threats to Validity}
\label{sec:threats}

\sectopic{Threats to construct validity} relate to our evaluation metrics and measurement pipeline. We report Precision, Recall, F1, FPR, the Matthews Correlation Coefficient (MCC), and AUPRC, and treat recall as the primary operational metric because a missed malicious skill file (false negative) carries a higher security cost than a blocked benign file, while FPR captures the deployability constraint; F1 is reported only for conventional comparison to prior work and is not the optimization target. 
Because SkillsBench is heavily imbalanced (9.1\% malicious), replicating a real world scenario where the vast majority of skill files are benign, we report MCC and AUPRC as our imbalance-robust summary metrics, MCC accounts for all four confusion-matrix cells and AUPRC is threshold-independent, so that the headline comparison does not hinge on F1, which can be inflated under class imbalance. 
Token savings are measured with tiktoken \texttt{o200k\_base} over the exact payloads the deployed system sends (preamble, skill header, and snippet windows) rather than simplified estimates, so the reported reductions match production behavior; nonetheless, our reliance on an LLM-as-a-judge for the final label could introduce measurement error relative to a purely manual review. 

\sectopic{Threats to internal validity} concern factors within the study that could influence the reported outcomes. The gpt-5.4-mini judge is nondeterministic, and results may shift across API versions or sampling settings. To account for this, every SkillsBench detection number we report for \toolname{} and SkillScanner+LLM is the mean of three independent runs (per-run std for the $ms{=}20$ default: F1 0.013, AUPRC 0.016, MCC 0.019, FPR 0.59pp); because the regex prefilter is deterministic (540 escalations per run), this variance is confined to the LLM judge, and the static baselines (ClawVet, SkillScanner) are invariant across runs. The three runs span a three-week window and may straddle a provider-side model update (per-run false positives 7/24/20); we report their mean, not any single run. The context window ($cw{=}500$) and confidence threshold ($\theta{=}0.70$) were pre-specified on a held-out development set. The snippet budget was set to $ms{=}20$ because it dominates the smaller $ms{=}5$ budget on that same development set (recall 0.972 vs.\ 0.928 at precision 1.0), so the choice is justified without consulting benchmark labels; nonetheless, we cannot rule out that a still different configuration would shift the accuracy--cost trade-off. 

\sectopic{Threats to external validity} concern the generalizability of our findings. The prefilter patterns are derived from the MITRE ATT\&CK taxonomy which is the most comprehensive pattern collection we could find, but cannot guarantee to cover all known security patterns. The benchmark's attack vocabulary may also not represent all real-world payloads or advanced obfuscation tactics, as evidenced by the residual encoding-based false negatives (RSA modular arithmetic, zero-width characters) that expose a concrete coverage gap. A further limitation is that SkillsBench's 150 malicious files are hand-crafted attack exemplars rather than captured in-the-wild samples, so measured recall may not transfer to organically evolving campaigns. All results use gpt-5.4-mini (OpenAI); recall on the LLM path may differ for other backends (e.g., claude-haiku-4-5 or local Ollama models), which we do not evaluate. Finally, the ClawHavoc campaign~\cite{clawhavoc2026} and other registry-scale incidents were not available as benchmark datasets, so performance on live adversarial skill packages, particularly novel techniques crafted to evade the current ruleset, may differ from our benchmark results.

\section{Conclusion}
\label{sec:conclusion}

In this paper, we present \toolname{}, a low-cost, deployable security proxy that screens MCP skill files before they reach an AI coding agent, closing the gap of the agent skills security challenge. \toolname{} pairs a 530-pattern regex prefilter with an LLM judge that sees only the flagged snippet windows: benign files bypass the model entirely, and flagged files are classified from their risky regions rather than their full content. On the SkillsBench benchmark (1{,}650 skills, 9.1\% malicious; 150 hand-crafted attack exemplars), \toolname{} achieves F1$=$0.817, recall$=$0.769, and FPR$=$1.13\% (MCC$=$0.803, AUPRC$=$0.830; mean of three runs), outperforming ClawVet and SkillScanner by 5--6$\times$ on threshold-independent AUPRC while reducing LLM input tokens by 77\% and screening each skill in ${\sim}818$ms, 7.7$\times$ faster than escalating every file to the same judge. These results show that targeted, snippet-level screening makes inline malicious-skill detection both accurate and inexpensive enough to run at install time. We hope \toolname{} serves as a practical first line of defense as the agent skill ecosystem, and its attack surface, continues to grow.

\balance
\bibliographystyle{IEEEtran}
\bibliography{references}

\end{document}